\documentclass[a4paper]{jpconf}

\newcommand{\apj}{Astrophys. J.}

\newcommand{\prd}{Phys. Rev. D}
\newcommand{\prl}{Phys. Rev. Lett.}

\usepackage{iopams,float}

\usepackage{epsfig,graphicx} %include figure files
\usepackage{amsmath, amsfonts,amssymb}
\usepackage{bm} %include bold math: \bm{} creates bold letters in math mode

\usepackage[linktocpage]{hyperref}
\usepackage[caption=false]{subfig}
\usepackage[usenames]{color}
\usepackage{multirow}

\def\be{\begin{equation}}
\def\ee{\end{equation}}
\def\bea{\begin{eqnarray}}
\def\eea{\end{eqnarray}}
\newcommand{\beq}{\begin{eqnarray}}
\newcommand{\eeq}{\end{eqnarray}}
\def\nn{\nonumber}

\begin{document}
%%%%%%%%%%%%%%%%%%%%%%%%%%%%%%%%%%%%%%%%%%%%%%%%%%%%%%%%%%%%%%%%%%%%%%%%%%%%%%%%%
\title{Environmental Effects for Gravitational-wave Astrophysics}
%%%%%%%%%%%%%%%%%%%%%%%%%%%%%%%%%%%%%%%%%%%%%%%%%%%%%%%%%%%%%%%%%%%%%%%%%%%%%%%%%

\author{Enrico Barausse$^{1,2}$, Vitor Cardoso$^{3,4}$ and Paolo Pani$^{3,5}$}
\address{$^{1}$CNRS, UMR 7095, Institut d'Astrophysique de Paris, 98bis Bd Arago, 75014 Paris, France}
\address{$^{2}$Sorbonne Universit\'es, UPMC Univ Paris 06, UMR 7095, 98bis Bd Arago, 75014 Paris, France}
\address{$^{3}$CENTRA, Departamento de F\'{\i}sica, Instituto Superior T\'ecnico, Universidade de Lisboa, Avenida Rovisco Pais 1, 1049 Lisboa, Portugal.}
\address{$^{4}$Perimeter Institute for Theoretical Physics, Waterloo, Ontario N2L 2Y5, Canada.}
\address{$^{5}$Dipartimento di Fisica, ``Sapienza'' Universit\`a di Roma, P.A. Moro 5, 00185, Roma, Italy}

\begin{abstract} 
The upcoming detection of gravitational waves by terrestrial interferometers will usher
in the era of gravitational-wave astronomy.
This will be particularly true when space-based detectors will come of age and
measure the mass and spin of massive black holes with exquisite precision and
up to very high redshifts, thus allowing for better understanding
of the symbiotic evolution of black holes with galaxies, and for high-precision tests of General Relativity
in strong-field, highly dynamical regimes.
Such ambitious goals require that astrophysical environmental pollution of gravitational-wave signals be
constrained to negligible levels, so that neither detection nor estimation of the source parameters are significantly affected.
Here, we consider the main sources for space-based detectors --         
the inspiral, merger and ringdown of massive black-hole binaries and extreme mass-ratio inspirals~--
and account for various effects on their gravitational waveforms, including electromagnetic fields, cosmological evolution,
accretion disks, dark matter, ``firewalls'' and possible deviations from General Relativity.             
We discover that the black-hole quasinormal modes are sharply different in the presence of matter, but the ringdown signal observed by interferometers is typically unaffected.
The effect of accretion disks and dark matter  depends critically on their geometry and density profile, but is
negligible for most sources, except for few special extreme mass-ratio inspirals.
Electromagnetic fields and cosmological effects are always negligible.
We finally explore the implications of our findings for proposed tests of General Relativity with gravitational waves, and
conclude that environmental effects will not prevent the                                     
development of precision gravitational-wave astronomy.
\end{abstract}

%%%%%%%%%%%%%%%%%%%%%%%%%%%%%%%%%%%%%%%
\section{Introduction}
%%%%%%%%%%%%%%%%%%%%%%%%%%%%%%%%%%%%%%%
Today, we have convincing indirect evidence from binary pulsars~\cite{HT1} for the existence of gravitational waves (GWs), which are a generic prediction of General Relativity (GR) and other relativistic theories of gravity. The ground-based GW detectors LIGO and Virgo are currently being updated to advanced configurations~\cite{Harry:2010zz,virgo} expected to achieve sensitivities
sufficient for detecting signals from binaries of stellar-mass black holes (BHs) and/or neutron stars within the end of this decade. Detection of these signals will allow measuring the masses and spins of the binary components with accuracies comparable to current X-ray probes~\cite{Vitale:2014mka}. On the same 
timescale, pulsar-timing arrays will target signals from widely separated massive-BH binaries~\cite{pta}, and on a longer timescale
spaced-based detectors will detect these systems at smaller separations, including the binary's merger and ringdown phases.  
Also, ESA has recently selected GWs as the science theme for its L3 mission with launch slot 2034. One possible mission that
would explore this science theme is given by the space-based detector eLISA~\cite{Seoane:2013qna}, whose  
``Pathfinder'' mission will be launched in 2015~\cite{pathfinder}.
Detectors such as eLISA will estimate the source parameters, and in particular the masses and spins of massive BHs, to within
 fractions of a percent and up to $z\sim 10-15$~\cite{Seoane:2013qna}, which will permit testing models for the 
symbiotic coevolution of massive BHs and their
host galaxies [see e.g. Refs.~\cite{Sesana:2010wy,Gair:2010bx,berti_volonteri,Barausse:2012fy,Sesana:2014bea}]. Furthermore, GW detectors will allow for precision tests of GR in the
currently unexplored highly dynamical, strong-field regime~\cite{Berti:2009kk,Gair:2012nm,Yunes:2013dva}.

Estimates of the accuracy of GW detectors in measuring the source parameters usually do not account for the realistic astrophysical environments
surrounding the sources  --~such
as electromagnetic fields, accretion disks and dark matter (DM)~-- 
based on the expectation/hope that their effect will be negligible.
However, a careful examination is needed to assess the environment's impact on GW observables, so as to determine
whether precision GW physics is possible at all: unmodeled
deviations (due to environmental effects) from the pure ``vacuum'' gravitational waveforms predicted
by GR may degrade the signal-to-noise ratio and the parameter estimation accuracy,
potentially jeopardizing tests of gravity theories and astrophysical models.
On the other hand, if these effects are non-negligible and can be modeled, they may provide important information about the environments of GW sources.

This article examines the impact 
of  environmental effects  on the most powerful source of GWs, namely the inspiral, merger and ringdown
of BH binaries, both with comparable and extreme mass-ratios. Our analysis follows that of Ref.~\cite{longer}, but we focus
here in particular on the implications for GW astrophysics with an eLISA-like mission.

%%%%%%%%%%%%%%%%%%%%%%%%%%%%%%%%%%%%%%%%%%%%%%%%%%%%%%%%%%%%%%%%%%%%%%%%%%
\section{Matter effects in  compact-object binaries}\label{sec:introdirty}
%%%%%%%%%%%%%%%%%%%%%%%%%%%%%%%%%%%%%%%%%%%%%%%%%%%%%%%%%%%%%%%%%%%%%%%%%%

The best understood environmental effect in compact-object binaries (including BH binaries) is the possible presence of a gaseous accretion disk.
During the binary's inspiral, the disk affects the orbital evolution in various ways. The masses and spins 
of the compact objects change under \emph{accretion} of gas~\cite{Barausse:2007dy,Macedo:2013qea}. 
Furthermore,  matter exerts a \emph{gravitational pull} on the binary's components, 
modifying their trajectories~\cite{Barausse:2006vt,Macedo:2013qea}. 
Finally, the gravitational interaction of the compact objects with their own 
wake in a gaseous medium produces \emph{dynamical friction}~\cite{Barausse:2007dy,Barausse:2007ph,Macedo:2013qea} and can give rise to \emph{planetary migration}~\cite{Yunes:2011ws,Kocsis:2011dr}.
All these effects impact the binary's orbital evolution, and therefore the gravitational waveforms. 

Other effects potentially affecting a binary's motion and  GW emission 
are the cosmological expansion and strong electromagnetic fields.
The former effects are expected to be small
and comparable to those of the cosmological constant, $\Lambda \sim 10^{-52}{\rm m}^{-2}$. 
As for the latter, astrophysical BHs 
are believed to be almost neutral because of quantum discharge effects~\cite{Gibbons:1975kk}, 
electron-positron pair production~\cite{1969ApJ...157..869G,1975ApJ...196...51R,Blandford:1977ds},
and charge neutralization by astrophysical plasmas.
Nevertheless, scenarios where BHs acquire an electric charge via various mechanisms,
typically related to the presence of a magnetic field, have been put forward to explain some BH-driven high-energy phenomena~\cite{1998ApJ...498..640P}.
Classical induction by an external magnetic field $B$ can produce a charge $Q$ satisfying~\cite{Wald:1974np}
\be
q\equiv\frac{Q}{\sqrt{G} M}\lesssim 2\times 10^{-6} \frac{M}{10^6 M_{\odot}}\frac{B}{\rm 10^{8} Gauss}\,.
\ee
The magnetic field of the massive BHs powering Active Galactic Nuclei (AGNs) is believed to be $\lesssim 10^3-10^5$ 
Gauss~\cite{2005ChJAS...5..347Z,2009A&A...507..171S,2013AstBu..68...14S}.
In the following,
to account for the uncertainties in the magnetic field measurements, we assume \textit{very} conservative upper limits, $B=10^8$ Gauss and $q=10^{-3}$.

DM might also affect the motion of BH binaries and their GW signal. 
Reference \cite{GS} showed 
that if a massive BH grows adiabatically, 
the DM density in its vicinity also increases and gives rise to a ``spike''.
However, these spikes were later found to be efficiently destroyed and diluted into ``cores'' by binaries of massive BHs with comparable masses~\cite{DMcoresBin}, thus making them less relevant for binaries living in galaxies that have recently experienced a major merger (i.e.~most Milky-way type galaxies). 
A similar flattening of these spikes is induced by DM scattering off stars~\cite{CoreScatter1,CoreScatter2} and by the possible off-center formation of the BH seeds at high redshift~\cite{ullio}.
The combination of these effects is expected to produce DM profiles with a shallow slope (rather than a spike) near BH binaries, with reference densities
$\rho_{\rm DM} \sim 10^2-10^3 M_\odot/{\rm pc}^3$ for binaries with mass ratio $1:1$ and $1:10$, respectively~\cite{DMcoresBin}.

In satellite galaxies that have never undergone mergers the situation might be different.
Reference \cite{SatelliteHalos} found that if massive BHs  grow from ``light'' seeds with mass $M_{\rm seed}\sim 10^2 M_{\odot}$ at $z\sim20$,
they may still be surrounded by DM spikes at $z\sim0$.
However, these spikes are unlikely to pose a problem for precision GW astronomy, because the BHs residing at 
their center would have mass $M_{\rm bh}\sim M_{\rm seed}$, and would thus emit outside the frequency band
of space-based detectors (such as eLISA). 
On the other hand, if massive BHs form from ``heavy'' seeds with mass $M_{\rm seed}\sim 10^5 M_{\odot}$ at $z\sim 15$,
Ref. \cite{SatelliteHalos} find that $\sim 100$ satellite galaxies containing BHs with mass $M_{\rm bh}\sim M_{\rm seed}$ surrounded by spikes would be present in
a Milky-Way type halo. These BHs might possibly produce GWs in eLISA's band, e.g. if they were to capture a stellar-mass BH and form
an extreme mass-ratio inspiral (EMRI), see also Ref. \cite{silk}.

Because of the various scenarios discussed above, we take our reference value for the DM density to be
\begin{equation}
\rho_{\rm DM}=10^3M_{\odot}/{\rm pc}^{3}=4\times 10^{4}{\rm GeV}/{\rm cm^3}\,,\label{rhoDM}
\end{equation}
for massive BH binaries with comparable masses and for EMRIs in Milky-way type galaxies,
but we also entertain the extreme possibility of higher DM densities (i.e.~spikes) of $\sim 10^{10}-10^{12} M_{\odot}/{\rm pc}^3$ in satellite galaxies.
%
%%%%%%%%%%%%%%%%%%%%%%%%%%%%%%%%%%%%%%%%%%%%%%%%%%%%%%%%%%%%%%%%%%%%%%%%%%%%%
\section{Environmental effects in the inspiral}\label{sec:inspiral}
%%%%%%%%%%%%%%%%%%%%%%%%%%%%%%%%%%%%%%%%%%%%%%%%%%%%%%%%%%%%%%%%%%%%%%%%%%%%%

Using the conservative reference values discussed above, we studied how realistic astrophysical environments
affect the inspiral phase of compact binaries,
during which the binary's separation shrinks adiabatically under GW emission. More specifically, we consider the gravitational pull of the matter surrounding the binary, the effect of gas accretion onto the binary's components, dynamical friction and planetary migration, BH charges induced by
external magnetic fields and cosmological expansion/acceleration effects.
We do not consider gravitational interactions with stars, which are known 
to be {\it (i)} unimportant for massive BH binaries in the eLISA band [see however 
\cite{pta2} for massive BH binaries detectable with pulsar-timing arrays] and {\it (ii)} 
only important for a few percent of the EMRIs detectable with eLISA~\cite{pau}, {\it if} stars have a cuspy distribution \textit{very close} to massive BHs [which is uncertain
observationally even for our Galaxy, c.f. Ref. \cite{schoedel}].
Similarly, gravitational interaction of an EMRI with a second nearby massive BH
(to within few tenths of a pc from the EMRI) may also produce a
detactable effect on eLISA waveforms~\cite{nico}. The fraction of EMRIs 
in such a situation is wildly uncertain, but probably on the order of a few percent
[see discussion in Ref.~\cite{nico}].

\begin{table}[ht]
\centering
\caption{\footnotesize Maximum environmental corrections to a typical EMRI's periastron shift ($\delta_{\rm per}$) and GW phase ($\delta_\varphi$) over a typical eLISA's mission duration of one year.
{$\delta_{\rm per}$ and $\delta_\varphi$ are respectively the relative and absolute corrections to vacuum values.
Dissipative effects such as GW radiation reaction, dynamical friction and hydrodynamic drag from accretion give negligible $\delta_{\rm per}$ and are thus not shown.
We consider two BHs with  masses $(10M_{\odot},M=10^6M_{\odot}$),  on a quasicircular inspiral
ending at the innermost stable circular orbit (ISCO) $r=6 G M/c^2$. 
The periastron shift is computed at $r=10 GM/c^2$.
Conservative environmental reference values are $q=10^{-3},\,B=10^{8}\,{\rm Gauss},\,\rho_3^{\rm DM}=\rho_{\rm DM}/(10^3 M_\odot/{\rm pc}^3)$. We assume a Shakura-Sunyaev disk model with viscosity parameter $\alpha=0.1$
and Eddington ratio $f_{\rm Edd}=10^{-4}$ ($f_{\rm Edd}=1$) for thick and thin disks, respectively. The scaling with the parameters can be found in Ref.~\cite{longer}, and in Ref.~\cite{Yunes:2011ws} for planetary migration. 
}
}
\vskip 0.3cm
\footnotesize
\begin{tabular}{cc|cc}
 \hline\hline
	    &  Correction             		  &$|\delta_{\rm per}|$	      & $|\delta_\varphi| [{\rm rads}]$ \\ \hline
\multirow{3}{*}{thin disks}	    &  planetary migration 		  &   ---		      & $10^4$ \\
 &  dyn. friction/accretion	  &   ---		      & $10^{2}$ \\
	    &  gravitational pull  	          & $10^{-8}$	      	      &	$10^{-3}$ \\ 
\hline
	    &  magnetic field        		  & $10^{-8}$	              & $10^{-4}$ \\
	    &  electric charge 		       		  & $10^{-7} $	              &	$10^{-2} $ \\
	    &  gas accretion   		  & $10^{-8}$	              &	$10^{-2}$ \\ 
	    & cosmological effects	&$10^{-31}$	              &	$10^{-26}$ \\ 
\hline
\multirow{2}{*}{thick disks} &  dyn. friction/accretion		  & ---	              &	$10^{-9}$ \\  
	    &  gravitational pull		  & $10^{-16}$	      &	$10^{-11}$ \\
\hline  
\multirow{3}{*}{DM}	    &  accretion               		  & ---		              &	$10^{-8}\rho_3^{\rm DM}$ \\
	    &  dynamical friction  		  & ---		              &	$10^{-14}\rho_3^{\rm DM}$\\
	    &  gravitational pull	          & $10^{-21}\rho_3^{\rm DM}$ &	$10^{-16}\rho_3^{\rm DM}$ \\
\hline\hline
\end{tabular}
\label{tab:periastron}
\vskip-0.1cm
\end{table}

Our main conclusions  are:

\begin{enumerate}
\item Environmental effects can be safely neglected for most compact-object binaries detectable with eLISA.
Table~\ref{tab:periastron} shows
the corrections to the periastron shift and GW phase
in a variety of environments, in the case of EMRIs, which are the most affected sources 
because they spend $\sim 10^5-10^6$ inspiral cycles in the frequency band of eLISA-like detectors (as opposed to $\sim 10^2$ cycles
for massive-BH binaries with comparable masses).
Even with our rather extreme reference values, matter corrections are 
at most marginally detectable and usually completely negligible. As an approximate criterion, for an (unmodeled) effect to have a measurable effect on eLISA waveforms,
it should introduce a dephasing $\delta_\varphi\gtrsim 0.1~{\rm rads}$ during the mission's expected lifetime of $1$ year.

\item An exception is given by EMRIs in geometrically-thin, radiatively-efficient accretion disks, where environmental 
effects are sufficiently large to affect estimates of the source parameters,
prevent accurate tests of GR, and possibly even jeopardize detection.
We found indeed that accretion, dynamical friction and planetary migration due to
a thin disk are typically more important than GW emission at separations $\gtrsim\sim 20-40$ gravitational radii, and
 are stronger than second-order gravitational self-force effects~\cite{2ndorderSF} at any separation~\cite{longer}.
The disk's gravitational pull is instead weaker.
This confirms and extends results by Refs.~\cite{Barausse:2007dy,Yunes:2011ws,Kocsis:2011dr,Macedo:2013qea}. 
However, EMRIs will be detectable with eLISA only at $z\lesssim 1$~\cite{Seoane:2013qna}, where most galactic nuclei
are quiescent rather than active. Because radiatively-efficient thin disks are mainly expected in AGNs, Ref.~\cite{longer} estimates that only a few percent
of the EMRIs detected by eLISA will be significantly affected. 
BHs in quiescent nuclei are expected to be surrounded by thick radiatively-inefficient disks, 
whose effect is typically sub-dominant even relative to second-order self-force corrections. 
Overall, our results thus confirm that EMRI detection and parameter estimation with eLISA should only be marginally affected by the environment. 
\item The early dynamics of  massive BH binaries with comparable masses in gas-rich environments (such as a dual AGN)
might be dominated by accretion and/or dynamical friction rather than by GW emission, 
from the time they enter the eLISA band down to $\sim 60-70$ gravitational radii. At smaller separations
(where most of the signal is produced) the dynamics is instead driven by GW radiation
reaction.
\item DM accretion and dynamical friction produce effects on BH binaries that are typically larger than the gravitational pull from
the DM halo. Nevertheless, all these effects are negligible for both detection and parameter estimation in the case of 
eLISA binaries, except (possibly) for EMRIs in satellite galaxies. As discussed in section~\ref{sec:introdirty}, DM spikes with densities $\rho\sim 10^9 - 10^{12} M_\odot /{\rm pc}^3$
may survive in such systems, in which case DM environmental effects on EMRIs (especially accretion onto the central massive BH) may be marginally comparable to GW emission.
\end{enumerate}
\vskip-0.8cm
\begin{figure}[h]
\begin{center}
\begin{tabular}{c}
\epsfig{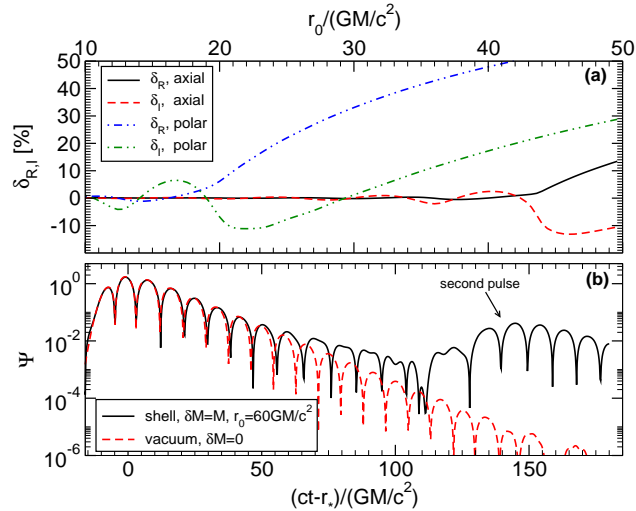}
\end{tabular}
\caption{\footnotesize (a) Deviations of the real and imaginary parts of the fundamental $l=2$ QNM frequencies for a BH with mass $M$ surrounded by a thin shell of mass $\delta M=0.01 M$ located at $r_0$. In the large-$r_0$ limit, $\delta_R$ and $\delta_I$ grow linearly. 
(b) Gravitational-axial waveform for the scattering of a gaussian profile off a thin shell - BH configuration with $r_0=60 GM/c^2$ and $\delta M=M$. The gaussian is initially placed close to the ISCO, with width $\sigma=4GM/c^2$. The effect of matter appears in the time domain as a second pulse at late times, but the isolated BH QNMs dominate most of the response.
\label{fig:ringdown}}
\end{center}
\end{figure}
%

%%%%%%%%%%%%%%%%%%%%%%%%%%%%%%%%%%%%%%%%%%%%%%%%%%%%%%%%%%%%%%%%%%%%%%
\section{Environmental effects in BH ringdown}\label{sec:ringdown}
% %%%%%%%%%%%%%%%%%%%%%%%%%%%%%%%%%%%%%%%%%%%%%%%%%%%%%%%%%%%%%%%%%%%%%%

In vacuum, the late-time signal after a massive BH merger is governed by the final BH's quasinormal modes (QNMs). 
This ``ringdown'' can be used to infer the remnant's mass and spin, and to test the no-hair theorem of GR with space-based detectors~\cite{Berti:2009kk}.
We have performed a detailed study of the ringdown emission of ``dirty BHs'', 
i.e.~BH geometries
not exactly described by the Schwarzschild metric
because containing matter~\cite{Visser:1992qh}. Our results can be summarized as follows [see Ref.~\cite{longer} for derivations and more details]:
\begin{enumerate}
\item The mode spectrum, as defined by the poles of the relevant Green's function in a \textit{frequency-domain analysis}, 
is extremely rich. For each mode of an isolated Schwarzschild BH we find an \emph{infinite} set of matter-driven modes, 
plus one  mode that parametrically corrects the isolated-BH mode (i.e.,~that reduces to it for vanishing matter stress-energy). 

\item For matter configurations localized far from the BH or very close to its horizon,   
the deviations from the isolated BH QNMs are arbitrarily large, even for very small matter densities. 
This surprising behavior is illustrated in the upper panel of Fig.~\ref{fig:ringdown},
for an infinitely thin shell of matter.
While a thin-shell geometry is 
astrophysically unrealistic, it is useful as a proof of concept~\cite{Leung:1999rh}. Our results show that this behavior is generic and appears for arbitrary composite BH-matter distributions~\cite{longer}.

\item An extensive search shows that although the spectrum of dirty BHs drastically differs from isolated ones,
the modes of the \emph{isolated} BH play a dominant role in a \textit{time-domain evolution} 
of the composite system. In other words, although the QNMs of isolated BHs are no longer poles of the Green's function, 
they nevertheless dominate the ringdown waveform at intermediate time scales. 
This is illustrated in the bottom panel of Fig.~\ref{fig:ringdown}, which shows the response of a thin shell -- BH system to a Gaussian perturbation.
After the Gaussian wavepacket ``hits'' the circular photon orbit, 
where isolated BH QNMs are known to reside~\cite{Cardoso:2008bp}, the \textit{composite} system rings down in its lowest {\it isolated} BH
QNMs. The modes of the composite system get excited
only at late times (when the wavepacket reaches the shell).

\item Our analysis confirms that the lowest QNMs of isolated BHs {\it can} be used to estimate the 
mass of massive BHs to  $0.1\%$ accuracy~\cite{Berti:2009kk}, in spite of our ignorance on the details of their astrophysical environments. 

\item The QNMs of dirty BHs are typically localized farther away from the object than in isolated geometries. 
During a BH merger, these modes are excited to low amplitudes and at very late times. Therefore, 
they typically play a subdominant role in the merger waveforms, but  will likely dominate over Price's power-law tails~\cite{Price:1971fb}. 
In principle, these modes might be excited to large amplitudes during a binary's inspiral, 
thus providing important information on the matter distribution, possibly even close to the horizon.

Consider for instance Planck-density ``firewalls'' [one possible consequence of quantum effects near the horizon~\cite{Almheiri:2012rt}, see also Ref.~\cite{Braunstein:2009my} for a similar earlier proposal], which we model by a spherical shell of mass $\delta M$ located at $r_0\sim 2M+\ell_P$, where $\ell_P\sim 1.6\times 10^{-35}{\rm m}$ is the Planck length. We estimate the (relative) changes in the fundamental QNM frequency to be given by 
\begin{equation}
\delta_R\sim -2\left[1-0.01\log\left(\frac{10^6M_\odot}{M}\right)\right] \frac{\delta M}{10^{-2} M} \,. \label{deltaRfirewall}
\end{equation}
This suggests that a firewall of mass as low as $\delta M\sim 10^{-4}M$ introduces a correction of the order of a few percent in the BH ringdown frequencies. Such corrections might be observable with Advanced LIGO/Virgo or eLISA, if the modes are excited to appreciable amplitudes. 
We stress, however, that the correction to the QNM frequencies depends  (linearly) on the total mass of the firewall, which is still subject of debate~\cite{Abramowicz:2013dla,Israel:2014eya}.
\end{enumerate}

\begin{table}[b]
\centering
\footnotesize
\caption{\footnotesize Upper limits on the environmental corrections to the BH QNMs.
{We define $\delta_{R,I}=1-\omega_{R,I}/\omega_{R,I}^{(0)}$, where $\omega_{R,I}$ is the real (imaginary)
part of the ringdown frequency in the presence of environmental effects, whereas $\omega_{R,I}^{(0)}$ 
is the same for an isolated BH with the same total mass.
Environment reference values are  as in Table~\ref{tab:periastron}. 
The spherical and ring-like matter distributions  have  mass $\delta M \sim 10^{-3}M$.
The scaling with the parameters is shown in Ref.~\cite{longer}.
}
}
\vskip 0.2cm
\begin{tabular}{c|cc}
 \hline\hline

Correction           	   	& $|\delta_R|[\%]$ 	&$|\delta_I|[\%]$   	\\
\hline
spherical near-horizon distribution
	&$0.05$	             	&$0.03$             	\\
ring at ISCO	 	       	&$0.01$	             	&$0.01$	            \\
electric charge 		               	&$10^{-5} $   	&$10^{-6}$    	\\
magnetic field           	&$10^{-8}$  	&$10^{-7}$	\\
gas accretion              	&$10^{-11} $  	&$10^{-11}$     \\
DM halos	 	       	&$10^{-21}\rho_3^{\rm DM}$ 	&$10^{-21}\rho_3^{\rm DM}$   	\\
cosmological effects 		&$10^{-32}$ 	&$10^{-32}$   	\\
\hline\hline
\end{tabular}
\vspace{0.2cm}
\label{bstable}
\end{table}

In summary, matter drastically changes the frequency-domain QNM spectrum, but does not prevent GW observatories from detecting BH ringdowns in realistic environments using isolated-BH templates. In very optimistic scenarios, detections could even be used to extract the mass distribution parameters and investigate accretion disks, DM halos or even firewalls~\cite{longer}. These statements are supported by Table~\ref{bstable}, which shows an upper limit to the QNM corrections in a variety of astrophysical scenarios. 

%

%%%%%%%%%%%%%%%%%%%%%%%%%%%%%%%%%%%%%%%%%%%%%%%%%%%%%%%%%%%%%%%%%%%%%%%%%%
\section{Intrinsic limits to tests of gravity theories}\label{sec:modifiedgravity}
%%%%%%%%%%%%%%%%%%%%%%%%%%%%%%%%%%%%%%%%%%%%%%%%%%%%%%%%%%%%%%%%%%%%%%%

Quantifying environmental effects around compact binaries is also important for testing GR.
So far, GR has been tested in situations 
involving weak gravitational fields and mildly-relativistic
velocities~\cite{Will,Will:2005va}.
GW observations of relativistic compact binaries will provide the possibility of testing GR in the unexplored strong-field, highly dynamical regime, i.e.~in systems with characteristic velocities $v\sim c$
and gravitational potentials/curvatures orders of magnitude larger than those probed so far~\cite{Psaltis:2008gka,Gair:2012nm,Yunes:2013dva}.
For these tests it is essential to understand the effect of matter, in order
to avoid mistaking environmental effects for deviations from GR. More specifically, matter effects
will provide an intrinsic limit to the precision of GR tests. If the gravity-theory modifications
introduce effects smaller than the environmental ones, the former will be extremely hard to detect, unless the environment is precisely modeled.
For instance, from the analysis of the previous section it is clear that testing GR will be extremely difficult with EMRIs in thin-disk environments, where planetary migration, dynamical friction and accretion are as or more important than GW emission.

\begin{table*}[b]
\centering
\caption{\footnotesize Intrinsic lower constraints on some modified gravity theories due to environmental effects in the orbital decay rate of a binary inspiral.
{We consider Brans-Dicke (BD) theory, Einstein-Dilaton-Gauss-Bonnet (EDGB) gravity, Dynamical Chern-Simons (DCS) gravity, Einstein-\ae ther theory (\AE) and Ho\v rava gravity. We define $v_3=v/(10^{-3}c)$, $\rho_3^{\rm DM}=\rho_{0}/(10^3 M_\odot/{\rm pc}^3)$, $\rho_2^{\rm disk}=\rho_0/(10^2 {\rm kg}/{\rm m}^3)$, $M_{10}=M_T/(10 M_\odot)$, $B_{8}=B/(10^{8}{\rm Gauss})$, $q_{3}=Q/(10^{-3}M)$, $R_{10}=R/(10 GM/c^2)$, $R_{\rm DM}=R/(7\times 10^6 GM/c^2)$, where $M_T$ is the total mass of the binary, $\nu$ is the symmetric mass ratio, $v$ is the orbital velocity, $R$ and $\rho_0$ are the characteristic location and density of the matter distributions. The dimensionless coupling constants $\omega_{\rm BD}$, $\zeta_3$, $\zeta_4$, ${\cal F}$ and the parameters $\gamma_{\hat \alpha}$, $S$, $\delta_m$ and $\beta_{\rm dCS}$ are defined as in Ref.~\cite{longer} and Ref.~\cite{Yunes:2013dva}.
For accretion disks we use a spherical distribution $\rho\sim \rho_0(R/r)^{\hat\alpha}$,
which can approximate either thick disks ($\rho_0\sim 10^{-10} {\rm kg/m}^3$, $\hat{\alpha}\sim 3/2$) or thin disks ($\rho_0\sim 10^2 {\rm kg/m}^3$, $\hat{\alpha}\sim 15/8$) .
The normalization coefficients ${\cal P}$ and ${\cal T}$ are defined in the last row and last column. }
}
\vskip 0.2cm
\resizebox{\columnwidth}{!}{
\footnotesize
\begin{tabular}{c|cccc|c}

\hline \hline
&\multicolumn{4}{c}{Intrinsic lower bound}  \\ 
\hline
\multirow{2}{*}{Theory}    & \multirow{2}{*}{magnetic fields} & Pull of DM profile  &  Pull of disk profile & \multirow{2}{*}{electric charge} & \multirow{2}{*}{coefficient ${{\cal T}}$}		 \\
    & 		    & $\rho\sim \rho_0(R/r)^{3/2}$ 	&  $\rho\sim \rho_0(R/r)^{\hat\alpha}$&  &	 \\
\hline 
BD		& $\omega_{\rm BD}^{-1}\gtrsim 10^{-6} {\cal P} {{\cal T}}    $ 	& $\omega_{\rm BD}^{-1}\gtrsim 10^{-19}{\cal P} {{\cal T}}$ & $\omega_{\rm BD}^{-1}\gtrsim 10^{-1-5{\hat \alpha}} {\cal P} {{\cal T}}$ & $\omega_{\rm BD}^{-1}\gtrsim 10^{-15} {\cal P} {{\cal T}} $ & $\left[\frac{0.1}{S}\right]^{2}$ \\
%%%
EDGB 		& $\zeta_3\gtrsim 10^{-12} {\cal P} {{\cal T}}$ & $\zeta_3\gtrsim 10^{-25} {\cal P} {{\cal T}}$ & $\zeta_3\gtrsim 10^{-7-5{\hat \alpha}} {\cal P} {{\cal T}}$ & $\zeta_3\gtrsim 10^{-21} {\cal P} {{\cal T}} $ & $\left[\frac{\nu}{0.1}\right]^{4}  \left[\frac{1}{\delta_m}\right]^{2}$\\
%%%
DCS		& $\zeta_4\gtrsim 10^{6} {\cal P} {{\cal T}}$ & $\zeta_4\gtrsim 10^{-7} {\cal P} {{\cal T}}$ & $\zeta_4\gtrsim 10^{-7-5{\hat \alpha}} {\cal P} {{\cal T}}$ & $\zeta_4\gtrsim 10^{-3} {\cal P} {{\cal T}}$ & $\left[\frac{\nu}{0.1}\right]^{2} v_3^{-6} \left[\frac{1}{\beta_{\rm dCS}}\right]$ \\
%%%%
\AE/Ho\v rava 	& ${\cal F}\gtrsim 10^{-9} {\cal P} {{\cal T}}$	& ${\cal F}\gtrsim 10^{-22}{\cal P} {{\cal T}}$	& ${\cal F}\gtrsim 10^{-4-5{\hat \alpha}} {\cal P} {{\cal T}}$	& ${\cal F}\gtrsim 10^{-18}{\cal P} {{\cal T}}$ & 1\\
\hline
coefficient ${\cal P}$		& $B_{8}^{2} M_{10}^{2} v_3^{-4}$ & $\rho_3^{\rm DM} M_{10}^{2} v_3^{-1} R_{\rm DM}^{3/2}  $ & $\gamma_{\hat \alpha} \rho_2^{\rm disk}M_{10}^{2} v_3^{2{\hat \alpha}-4} R_{10}^{{\hat \alpha}} $ & $q_3^2 v_3^{4}$ \\
\hline \hline
\end{tabular}
}
\label{tab:dotPmodified_dirtiness}
\end{table*}

In this section we therefore focus on systems and environmental effects that are \textit{smaller} than GW emission,
and compare them to possible deviations from GR, considering both the ``conservative effects'' due to the modified spacetime geometry,
and the ``dissipative effects'' caused by the modified GW emission. We perform this analysis for a generic gravity theory,
%%%%%%
%
\begin{eqnarray}
 \!S&=&\frac{c^4}{16\pi{\cal G}}\!\int\!\! {dx^4\!\sqrt{-g}}\left[R\!+\!\partial^2\bm{\Psi}\!+\!\sum_ia_i U_i(\bm{\Psi},\bm{g},\partial\bm{\Psi},\partial\bm{g},...)\!\right]\nn\\
 &+&S_m^{(0)}[\bm{\Psi}_m,g_{\mu\nu}]+\sum_i b_i S_{m,i}^{(1)}[\bm{\Psi}_m,\bm{\Psi},\bm{g},...]\,,\label{modifiedGR}
\end{eqnarray}
where $\bm{\Psi}$ is a generic field encoding the corrections to GR, ${\cal G}$ is an effective gravitational coupling (not necessarily matching
the value of $G$ measured by a Cavendish-type experiment), $\bm{\Psi}_m$ schematically represents the matter fields, $U_i$ are nonminimal interaction terms, and we have linearized the matter action $S_m$. The couplings $a_i$ and $b_i$ parametrize deviations from GR (i.e.~the theory reduces to GR for $a_i=b_i=0$) and have different physical dimensions, depending on the theory. 
All corrections to GR will be proportional to some power of $a_i$ and $b_i$. Although approximate, our analysis is largely theory-independent, as the action \eqref{modifiedGR} includes
most modifications of GR proposed in the literature [e.g. scalar-tensor theories, Chern-Simons gravity, khronometric gravity, Einstein-\ae ther theory, etc; see Ref.~\cite{longer} for details]. 

In Table~\ref{tab:dotPmodified_dirtiness} we present the intrinsic lower constraints
that can be placed by measuring the orbital decay rate of a binary inspiral
in some widely studied modified gravity theories.
These particular cases can be mapped to the generic parametrization~\eqref{modifiedGR} by identifying each theory's coupling constants to suitable powers 
of $a_i$ or $b_i$. As can be seen, these intrinsic limits are much less stringent than current observational bounds.

%%%%%%%%%%%%%%%%%%%%%%%%%%%%%%%%%%%%%%%%%%%%%%%%%%%%%%%%%%%%%%%%%%%%%%
\section{Conclusions}\label{sec:conclusions}
%%%%%%%%%%%%%%%%%%%%%%%%%%%%%%%%%%%%%%%%%%%%%%%%%%%%%%%%%%%%%%%%%%%%%%
We have quantified the impact of realistic astrophysical environments on GW signals from BH binaries,
including the effect of electromagnetic fields, cosmological evolution, accretion disks and DM.
Our analysis shows that GW astronomy has the potential to become a precision discipline, because 
environmental effects are typically too small to affect the detection of GW signals and the estimation of the source's
parameters. The few and rather extreme cases in which environmental effects might leave a detectable imprint
should be seen as an opportunity, i.e.~given a sufficiently sensitive detector and adequate modeling of these effects, 
GW astrophysics can be used to obtain information about the density and velocity of the matter surrounding GW sources.

For example, the DM density profile in galactic nuclei is still poorly understood, e.g. DM spikes may grow around the massive BHs 
dwelling at the center of galaxies~\cite{GS}. Our analysis shows that if these DM spikes survive till $z\sim 0$ 
(as may be the case in satellite galaxies~\cite{SatelliteHalos}), their effect would be detectable by eLISA in EMRIs, and GW observations could be used to constrain DM profiles near massive BHs (c.f. also Ref.~\cite{silk}).
Likewise, the corrections due to accretion disks depend on their geometry and on the accretion rate. EMRIs with exceptionally large SNR could 
be used to constrain models of accretion onto massive BHs. These intriguing possibilities requires more sophisticated modeling, which is beyond the scope of our analysis.
(See however Ref.~\cite{Barausse:2007dy,Macedo:2013qea,Barausse:2006vt,Barausse:2007ph,Yunes:2011ws,Kocsis:2011dr} for some work in this direction.)

A quantitative analysis of environmental effects is also vital to allow tests of GR to be performed with GW observations, 
without mistaking the effect of astrophysical matter for a breakdown of GR. Our results therefore
yield intrinsic lower bounds on the magnitude of the deviations from GR that can be tested with space-based GW detectors.
Although admittedly approximate, ours is a largely model- and theory-independent analysis. 
While sufficient for our purposes, more sophisticated modeling 
(e.g. including the effect of BH spins)
would be required to estimate environmental effects from real GW data when they become available.

Finally, our analysis has considered enviromental effects on a single source. If the effect under consideration is universal (as would be the
case for a modification of gravity), one may enhance eLISA's sensitivity to it by combining different sources. One might try to apply
a similar technique to matter effects such as those due to accretion disks, DM, etc. However, in that case the effects may be completely different in
different sources, and it is not guaranteed that correlating several sources will help detect them.

%%%%%%%%%%%%%%%%%%%%%%%%%%%%%%%%%%%%%%%%%%%%%%%%%%%%%%%%%%%%%%%%%%%%%%%%%%%%%%%%%%%%%%%%%%%%%%%%%%%%%%%%%%%%%%%%%%%%
\subsection*{Acknowledgments}
We thank Pau Amaro-Seoane, Leor Barack, Emanuele Berti, Marc Casals, Luciano Rezzolla, Joe Silk and Nico Yunes for useful comments.
This work was supported by the European Union through grants GALFORMBHS PCIG11-GA-321608, NRHEP 295189-IRSES, DyBHo--256667 ERC Starting Grant, aStronGR-IEF-298297,  AstroGRAphy-2013-623439 and by FCT-Portugal through projects CERN/FP/123593/2011, IF/00293/2013.
Research at Perimeter Institute is supported by the Government of Canada through 
Industry Canada and by the Province of Ontario through the Ministry of Economic Development 
\& Innovation.
Computations were performed on the ``Baltasar Sete-Sois'' cluster at IST,
XSEDE clusters SDSC Trestles and NICS Kraken through NSF Grant~No.~PHY-090003.
%%%%

\section*{References}

\end{document}